\documentclass[%
 reprint,
 amsmath,amssymb,
 aps,
]{revtex4-1}

\usepackage{epsfig,bm,epsf,graphics}
\usepackage{graphicx}
\usepackage{dcolumn}
\usepackage{bm}

\begin{document}

\preprint{APS/123-QED}

\title{Hexagonal-Close-Packed Lattice: Phase Transition and Spin Transport}

\author{Danh-Tai Hoang}%
 \email{danh-tai.hoang@u-cergy.fr}
\author{H. T. Diep}
\email{diep@u-cergy.fr, corresponding author}
 \affiliation{%
Laboratoire de Physique Th\'eorique et Mod\'elisation,
Universit\'e de Cergy-Pontoise, CNRS, UMR 8089\\
2, Avenue Adolphe Chauvin, 95302 Cergy-Pontoise Cedex, France.\\
 }%




\date{\today}

\begin{abstract}
We  study the  ground state (GS) and the phase transition in a hexagonal-close-packed lattice with both XY and Ising models by using extensive Monte Carlo simulation.  We suppose the in-plane interaction $J_1$ and inter-plane interaction $J_2$, both antiferromagnetic. The system is frustrated with two kinds of GS configuration below and above a critical value  of $\eta=J_1/J_2$ ($\eta_c$).   For the Ising case, one has $\eta_c=0.5$ which separates in-plane ferromagnetic and antiferromagnetic states, while for the XY case $\eta_c=1/3$ separates the collinear and non collinear spin configurations.  The phase transition is shown to be of first (second) order for $\eta> (<) \eta_c$.  The spin resistivity is calculated for the Ising case. It shows a rounded maximum at the magnetic transition in the second-order region, and a discontinuity in the first-order region of $\eta$.
\begin{description}
\item[PACS numbers: 05.70.Fh ; 75.10.Hk  ; 75.40.Mg ;75.47.-m  ]
\end{description}
\end{abstract}

\pacs{Valid PACS appear here}
\maketitle


\section{Introduction}

Frustrated spin systems have been subject of intensive investigations during the last 30 years \cite{Diep2005}.  These systems are very unstable due to the competition between antagonist interactions or due to a geometrical frustration such as in the antiferromagnetic (AF) triangular lattice.   A spin is said frustrated when it cannot find an orientation which "fully" satisfies all the interactions with its neighbors \cite{Toulouse,Villain}.  As a consequence of the frustration,  the ground state (GS) is very  highly degenerate.  In the Ising case the GS degeneracy is often infinite as in the triangular lattice,  the face-centered cubic (FCC) and hexagonal-close-packed (HCP) lattices, with AF interaction. In the case of vector spins, the GS is non collinear such as the 120-degree configuration in the XY and Heisenberg AF stacked triangular lattice (STL).      In two dimensions (2D), several frustrated systems with Ising spin model have been exactly solved \cite{Liebmann,Diep2005a}.  Among the most interesting models one can mention the frustrated generalized Kagome lattice \cite{Diep1991b} and the honeycomb lattice \cite{Diep1991a} where exotic features such as the existence of several phase transitions, the reentrance and the disorder lines have been exactly found by mapping these systems into vertex models \cite{Baxter}.
In three dimensions (3D), the situation is complicated. The renormalization group (RG) \cite{Wilson,Zinn}, which provided a good understanding of the nature of the phase transition in non frustrated systems,  encounters much of difficulties in application to frustrated systems.  Among the most studied subjects during the last 20 years, one can mention the nature of the phase transition in the XY and Heisenberg STL.  After a long debate \cite{Delamotte2005} on whether it is a second-,  or a first-order transition or it belongs to a new universality class, the controversy has recently ended with the conclusion of a first-order transition \cite{Ngo2008a,Ngo2008b}.

In this paper, we are interested in the HCP antiferromagnet with Ising and XY spin models.  Our purpose is to study its properties such as the ground state, the phase transition and the spin transport, in the case of anisotropic exchange interactions.  The isotropic nearest-neighbor (NN) AF interaction has been studied  for Ising \cite{Auerbach} and XY and Heisenberg spins \cite{Diep1992}.  These isotropic
cases have been  shown  to undergo
a phase transition of  first order, and the infinite
GS degeneracy is reduced to 6 at low temperatures \cite{Auerbach,Diep1992}. The effect of anisotropic interaction and anisotropy on the GS in the case of vector spins has been studied \cite{Larson}.

In section \ref{model}, we show our model and analyze the ground state.  Results of Monte Carlo (MC) simulations are shown  and discussed in section \ref{result}.
Concluding remarks are given in section \ref{conclu}.

\section{Model and Ground-State Analysis}\label{model}

\begin{figure}
\centering
\includegraphics[width=50mm,angle=0]{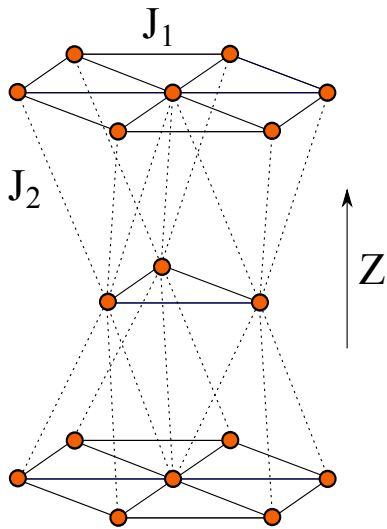}
\caption{HCP lattice.   The in-plane and inter-plane interactions are indicated by $J_1$ and $J_2$.} \label{hcp}
\end{figure}

We consider the HCP lattice shown in Fig. \ref{hcp}. The stacking direction is $z$.  The Hamiltonian is given by

\begin{equation}\label{HL}
{\cal H} = -\sum_{(i,j)}J_{i,j}\mathbf S_{i} \cdot \mathbf S_{j}
\end{equation}
where $\mathbf S_{i}$ is the spin at  lattice site $i$ and  $J_{ij}$ is the AF interaction between nearest-neighbors (NN). We suppose that $J_{ij}=J_1$ if the NN are on the $xy$ triangular plane, and $J_{ij}=J_2$ if the NN are on two adjacent planes (see Fig. \ref{hcp}).    The GS can be determined by the steepest-descent method.
This method is very simple \cite{Ngo2007} (i) we generate an initial configuration at random (ii) we calculate the local field created at a site by its neighbors using  (\ref{HL})  (iii) we align the spin of that site along the calculated local field  to minimize its energy  (iv) we go to another site and repeat until all sites are visited: we say we make one sweep (v) we  do a large number of sweeps per site until a good convergence is reached.

However, one can also minimize the interaction energy as shown below to calculate the GS configuration.  Since both interactions are AF (negative), for simplicity, we fix $J_2=-1$ and vary $J_1$.   The unit of energy is taken as $|J_2|$ and the temperature $T$ is in the unit of $|J_2|/k_B$ where $k_B$ is the Boltzmann constant.

Let us recall the GS in the case of isotropic interaction, namely $J_1=J_2$ \cite{Diep1992}.  For the HCP lattice, each spin is shared by
eight tetrahedra (four in the upper half-space and four in
the lower half-space along the z axis) and a NN bond is
shared by two tetrahedra. The GS spin configuration of
the system is formed by stacking neighboring tetrahedra.
In the GS, one has two pairs of antiparallel spins on each tetrahedron.  Their axes form an arbitrary angle
 $\alpha$.  The degeneracy is thus infinite (see Fig. 2a of Ref. \cite{Diep1992}). Of course, the periodic boundary conditions
will reduce a number of configurations, but the
degeneracy is still infinite. Of these GS's, one particular
family of configurations of interest for both XY and
Heisenberg cases is when $\alpha=0$. The GS is then collinear with two spins up and the other two down. The stacking sequence is then simplest: there are
three equivalent configurations since there are three ways
to choose the parallel spin pair in the original
tetrahedron.

We now examine the case where $J_1\neq J_2$.

\begin{figure}
\centering
\includegraphics[width=50mm,angle=0]{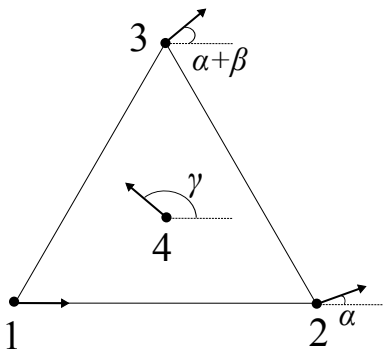}
\caption{Ground state in the  XY case.  The tetrahedron is projected on the $xy$ plane. The spins are numbered from 1 to 4.  See text for comments.} \label{GS}
\end{figure}

\begin{figure}
\centering
\includegraphics[width=50mm,angle=0]{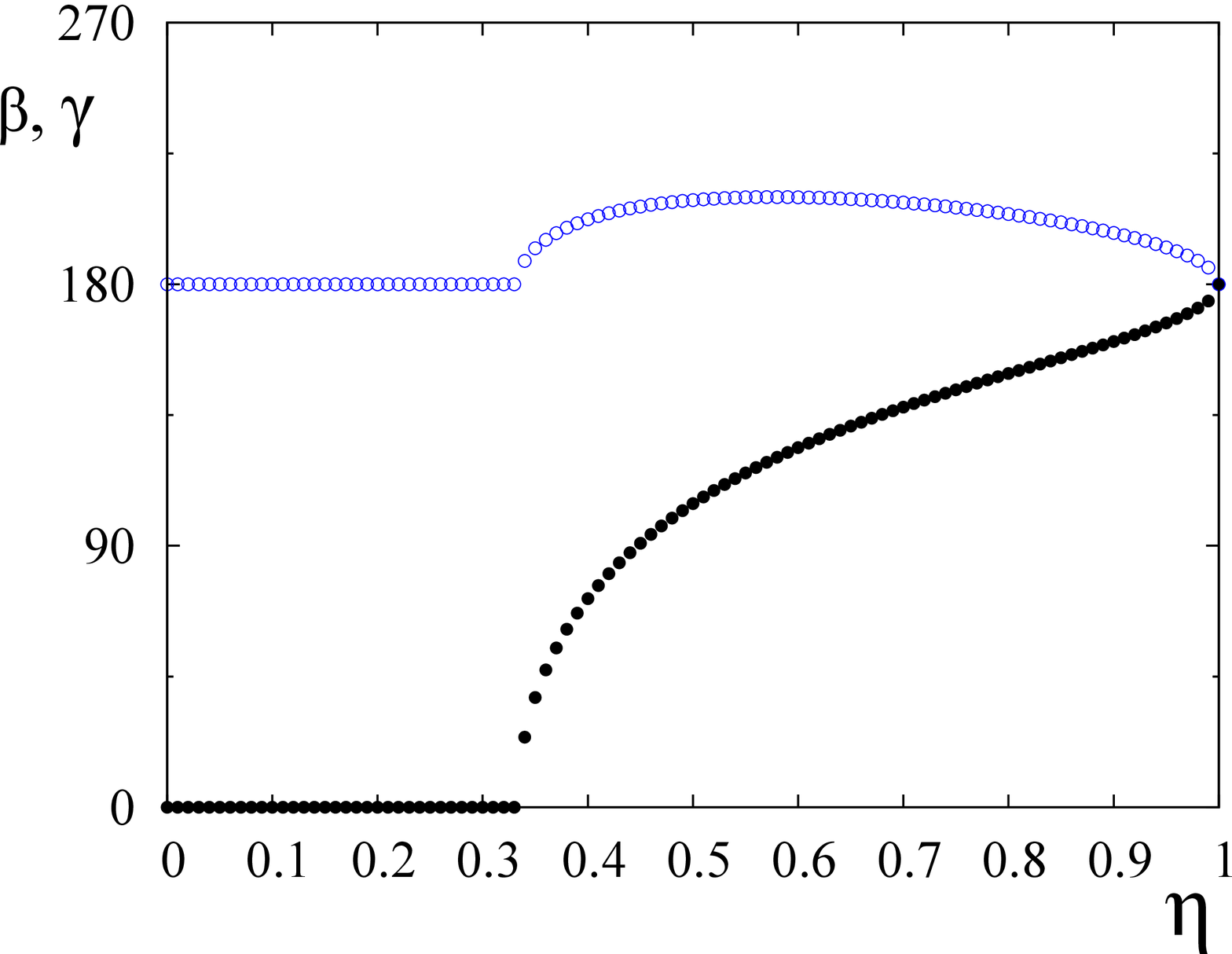}
\includegraphics[width=50mm,angle=0]{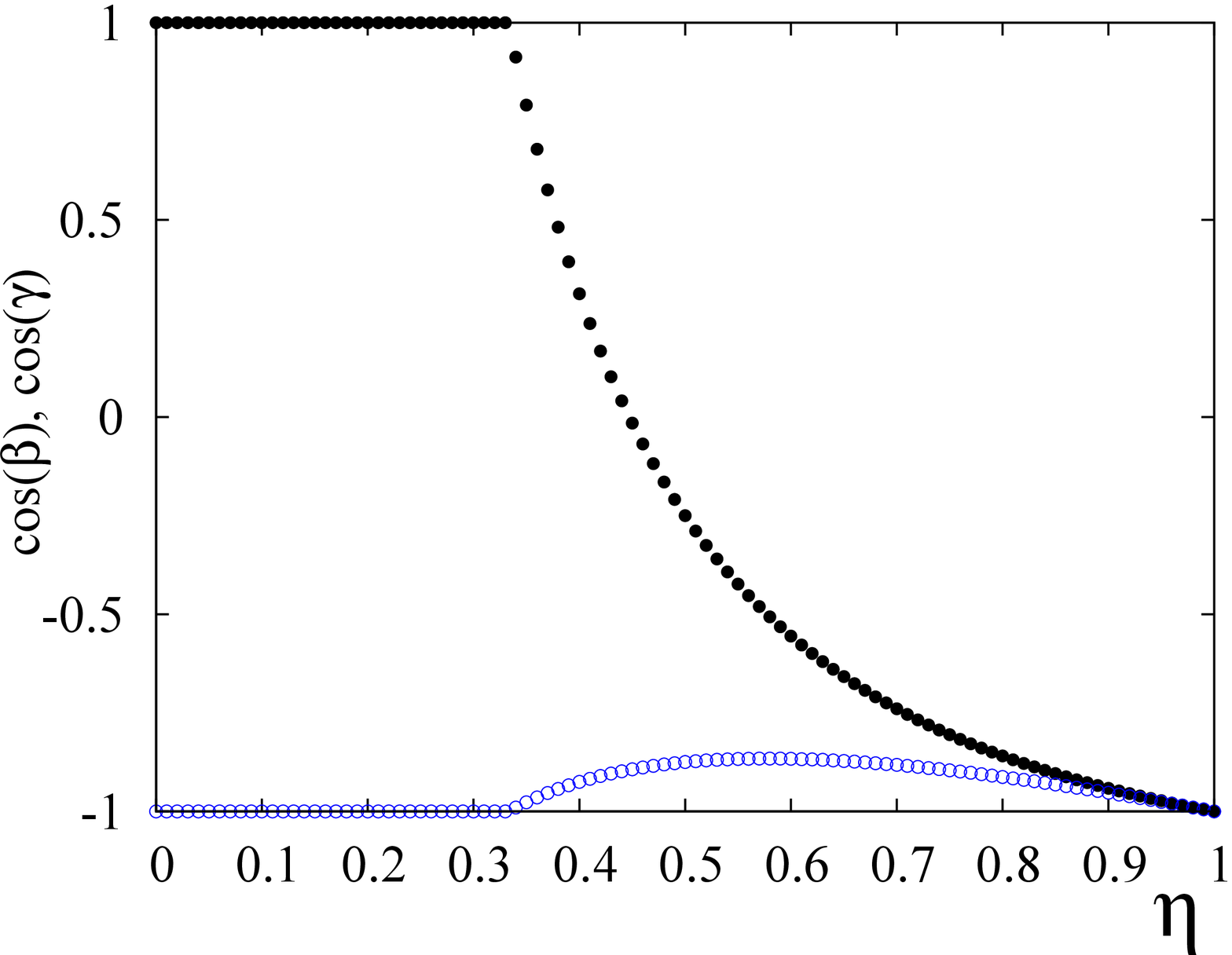}
\caption{Ground state in the XY case  (color online). The angles $\beta$ (black circles) and $\gamma$ (blue void circles) and their cosinus are shown as functions of $\eta=J_1/J_2$. Non collinear GS configurations occur in the region $1/3 \leq \eta \leq 1$ . See text for comments.} \label{angles}
\end{figure}

\begin{itemize}
\item Ising case:

The steepest descent method with varying $J_1$ ($J_2=-1$ gives two kinds of GS spin configuration: the first consists of $xy$ ferromagnetic planes stacked antiferromagnetically along the $z$ direction, while the second one is the stacking of $xy$ AF planes such that each tetrahedron has two up and two down spins.  The transition between the two configurations is determined as follows: one simply writes down the respective energies of a tetrahedron and compares them

\begin{eqnarray}
E_1&=&3(-J_1+J_2)\\
E_2&=&J_1+J_2
\end{eqnarray}
One sees that $ E_1<E_2$ when $J_1>0.5J_2$,  i.e.  $|J_1|<0.5 |J_2| $.
Thus the first configuration is more stable when $|J_1|<0.5 |J_2| $.

\item  XY case:

Consider one tetrahedron of the HCP lattice. The Hamiltonian for this cell is given by
\begin{eqnarray}
{\cal H}_c&=&-J_1(\mathbf  S_1\cdot \mathbf S_2+\mathbf S_2
\cdot \mathbf S_3+\mathbf S_3 \cdot \mathbf S_1)\nonumber \\
&&-J_2(\mathbf S_1 \cdot \mathbf S_4+\mathbf S_2\cdot \mathbf S_4
+\mathbf S_3\cdot\mathbf S_4)
\end{eqnarray}
Suppose that $|\mathbf S_i| = 1$, one has
\begin{eqnarray}
{\cal H}_c&=&-J_1 \left[ \cos\alpha+\cos\beta+\cos(\alpha+\beta)\right]\nonumber \\
&&-J_2 \left[\cos\gamma+\cos(\gamma-\alpha)+\cos (\gamma-\alpha-\beta)\right]
\end{eqnarray}
where the angles are defined in Fig. \ref{GS}.  The steepest descent method shows that while $\beta$ and $\alpha$ have unique values for a given $J_1/J_2$, $\alpha$ is arbitrary, just as in the case of isotropic interaction \cite{Diep1992} discussed above. To simplify the formulae, we take $\alpha=0$ in the following. The energy of the cell is written as
\begin{equation}
{\cal H}_c=-J_1 \left[1+2cos\beta\right]-J_2 \left[2\cos\gamma+\cos(\gamma-\beta)\right]
\end{equation}
The critical values of $\beta $ and $\gamma$ are determined from the relations
\begin{eqnarray}
\frac{\partial{\cal H}_c}{\partial\beta}&=&2J_1\sin\beta - 2J_2\sin(\gamma - \beta)=0\\
\frac{\partial{\cal H}_c}{\partial\gamma}&=&2J_2\sin\gamma + J_2\sin(\gamma - \beta)=0
\end{eqnarray}
We find the following solutions:

(i) $\beta=0$, $\gamma=0$,

(ii)$\beta=0$, $\gamma=\pi$,

(iii) $\beta=\pi$, $\gamma=0$,

(iv) $\beta=\pi$, $\gamma=\pi$, and

(v) $\cos\beta =\frac{1}{4(J_1/J_2)^2}-\frac{5}{4}$,  $\cos\gamma =-\frac{1+3(J_1/J_2)^2}{4(J_1/J_2)}$.

 By comparing the energy values at these solutions, we obtain the minimum energy with the last solution: one has
\begin{equation}
 {\cal H}_c=-\frac{1+3(J_1/J_2)^2}{2(J_1/J_2)}
 \end{equation}
  where $ \cos\beta$ and $\cos\gamma$ are given above.

Because $-1 \leq \cos\beta \leq 1 $ and $-1 \leq \cos\gamma \leq 1 $, the above solution is valid for $\frac{1}{3} \leq J_1/J_2 \leq 1 $.   We plot $\cos\beta$, $\cos\gamma$, $\beta$ and $\gamma$ in Fig. \ref{angles} where we observe that the non collinear GS configuration occurs in the interval $1/3 \leq \eta=J_1/J_2 \leq 1$.

\end{itemize}

.

\section{Phase Transition: Results}\label{result}

We consider  a sample size of $L\times L\times L_z$ where $L$ and $L_z$ vary from 12 to 36 but $L_z$ can be different from $L$ in order to detect the dependence of the GS on $J_1$.  The  exchange interaction $|J_2|=1$ is used as the unit of energy.  We use here MC simulations with a histogram technique to detect first-order transition. We    equilibrate the system and  average physical quantities over several millions of MC steps per spin.  The averaged energy $\langle U\rangle$ and the heat capacity $C_V$ are calculated by

\begin{eqnarray}
 \langle U\rangle&=&\langle {\cal H}\rangle\\
C_V&=&\frac{\langle U^2\rangle-\langle U\rangle^2}{k_BT^2}
\end{eqnarray}
where $\langle...\rangle$ indicates the thermal average taken over microscopic states at $T$.

The order parameter $M$  is defined from the sublattice magnetization by
\begin{equation}\label{M}
M=\sum_K |\sum_{i\in K} \mathbf S_i|
\end{equation}
where $\mathbf S_i$ belongs to the sublattice $K$. Note that there are at least four sublattices which define the ordering of the spins on the tetrahedra if a long-range order is observed.
The susceptibility is defined by
\begin{equation}\label{chi}
\chi=\frac{\langle M^2\rangle-\langle M\rangle^2}{k_BT}
\end{equation}

In MC simulations, we work at finite sizes, so for each size we have to determine the "pseudo" transition which corresponds in general to the maximum of the specific heat or of the susceptibility. The
maxima of these quantities need not to be at the same temperature. Only at the infinite size, they should coincide. The theory of finite-size scaling\cite{Hohenberg,Ferrenberg1,Ferrenberg2} permits to deduce properties of a system at its thermodynamic limit.  In order to check the first-order nature of the transition, we used the histogram technique which is very efficient in detecting weak first-order transitions and in calculating critical exponents of second-order transitions.\cite{Ferrenberg1,Ferrenberg2}  The main idea of this technique is to make an energy histogram at a temperature $T_0$ as close as possible to the transition temperature. Often, one has to try at several temperatures in the transition region.  Using this histogram in the formulae of statistical physics for canonical distribution, one obtains energy histograms in a range of temperature around $T_0$.   In second-order transitions, these histograms are gaussian. They allows us to calculate averages of physical quantities as well as critical exponents using the finite-size scaling.  In first-order transitions, the energy histogram shows a double-peak structure.

\subsection{Ising case}
\begin{figure}
\centering
\includegraphics[width=50mm,angle=0]{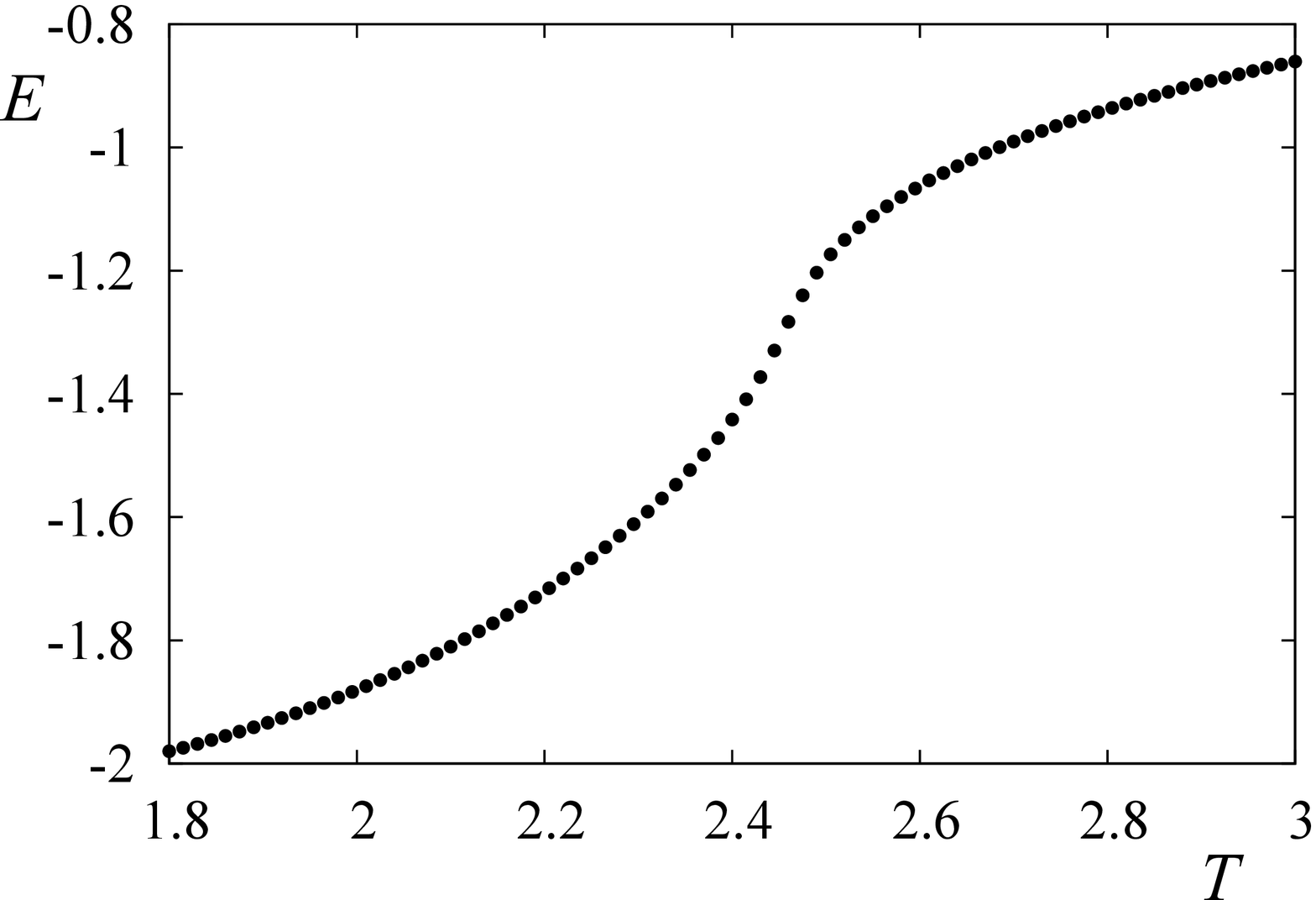}
\includegraphics[width=50mm,angle=0]{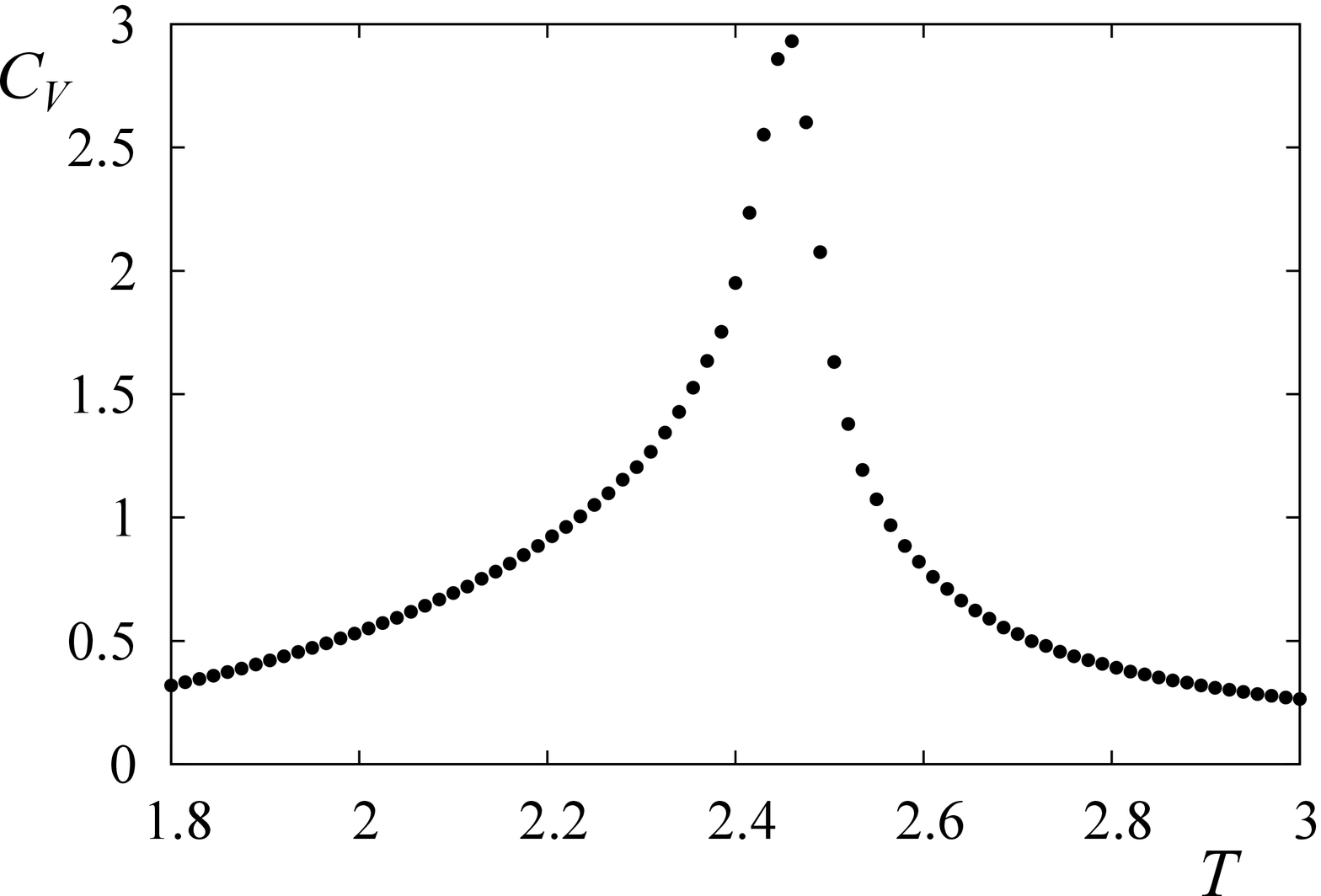}
\includegraphics[width=50mm,angle=0]{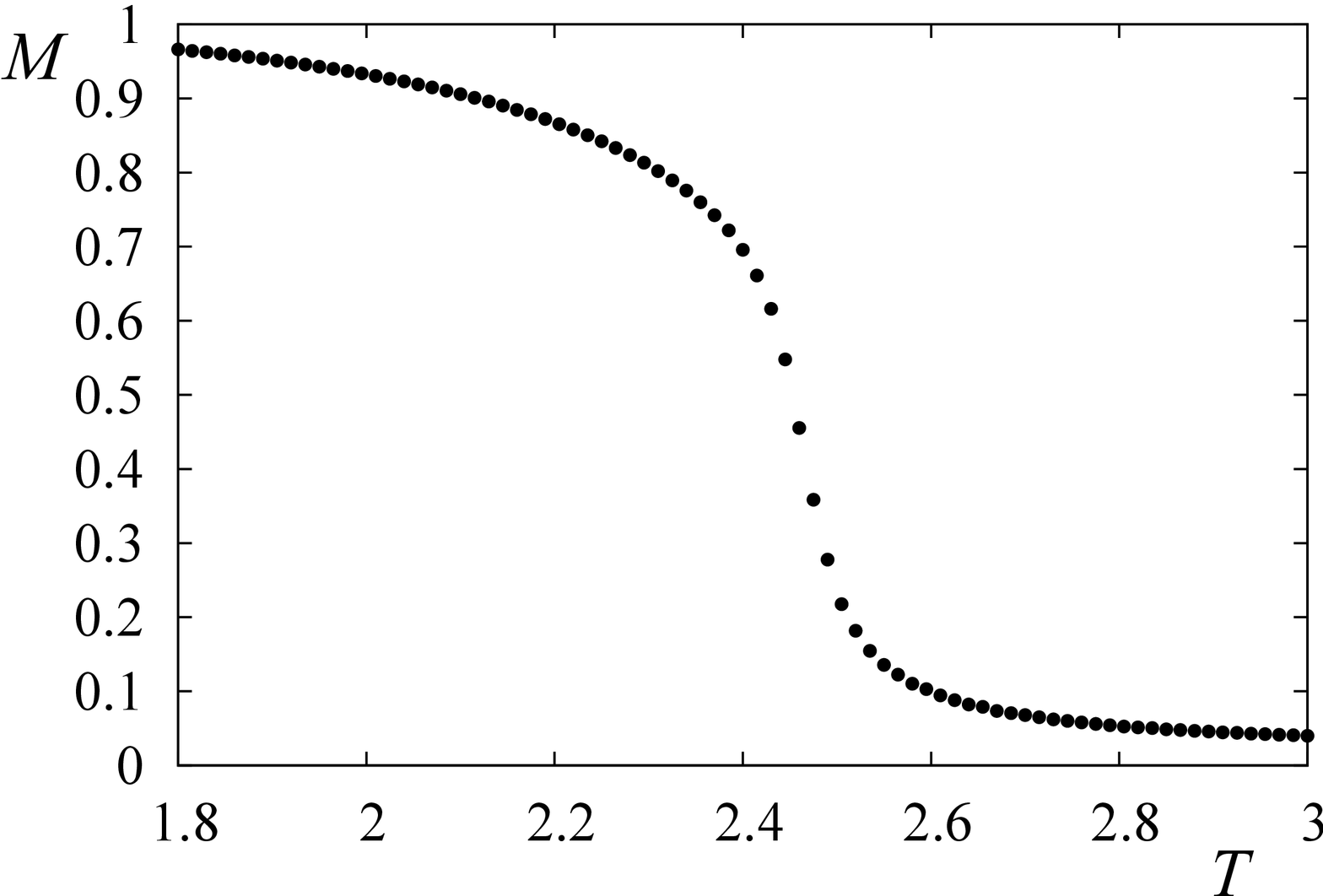}
\includegraphics[width=50mm,angle=0]{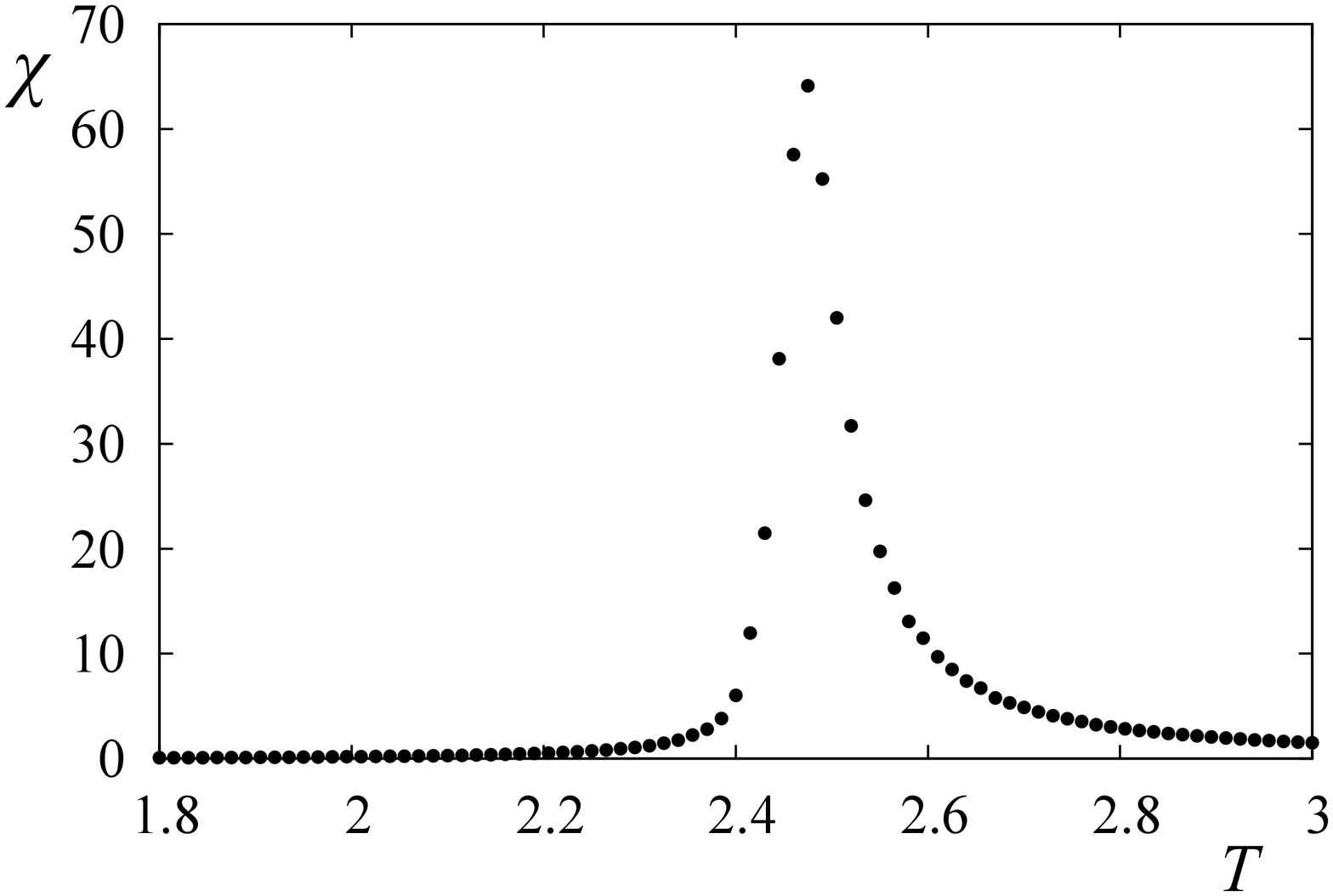}
\caption{Ising case:  Energy $E$, specific heat $C_V$, order parameter $M$ and susceptibility $\chi$ versus $T$ for $\eta=J_1/J_2=0.3$.   See text for comments.} \label{EMI03}
\end{figure}

\begin{figure}
\centering
\includegraphics[width=50mm,angle=0]{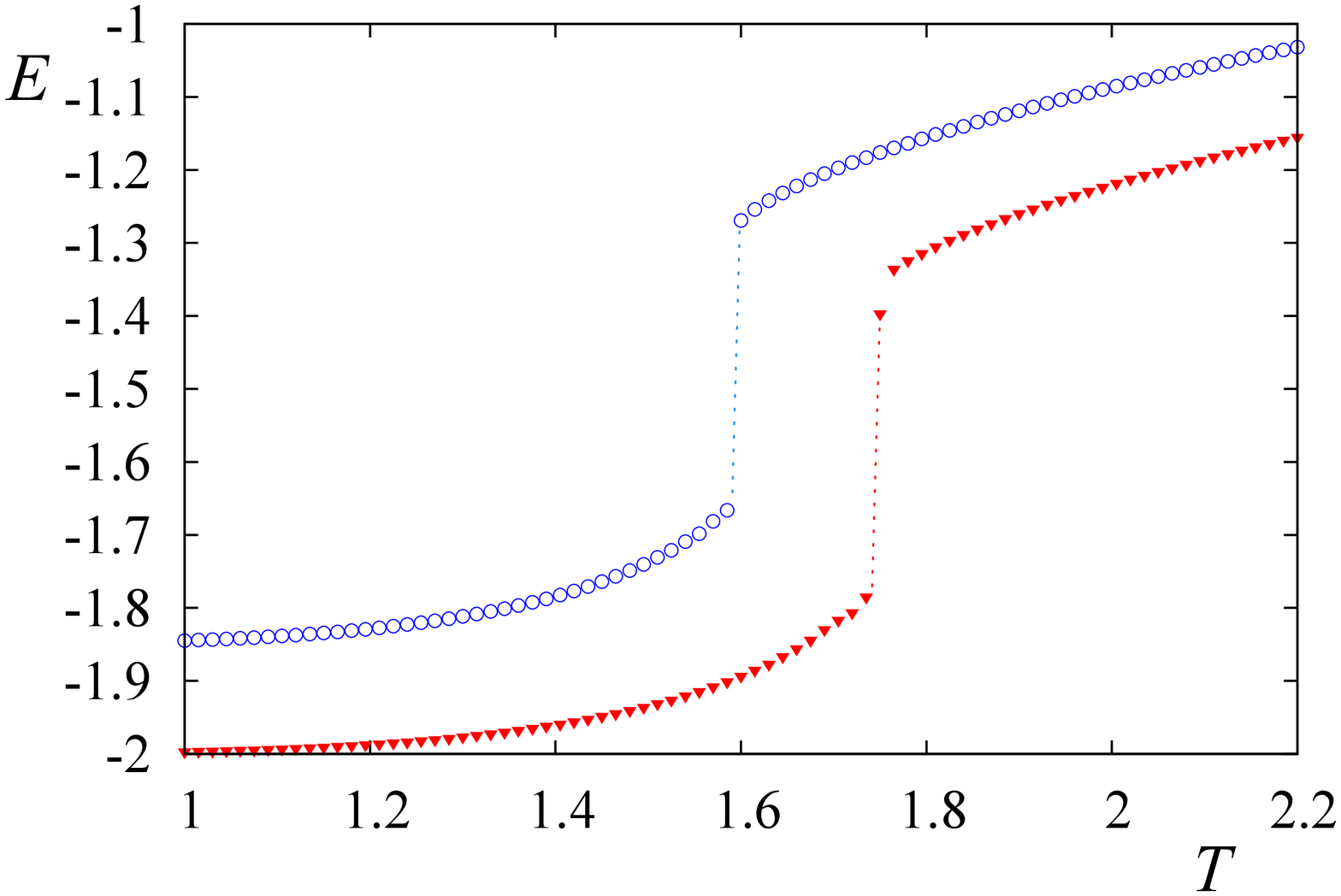}
\includegraphics[width=50mm,angle=0]{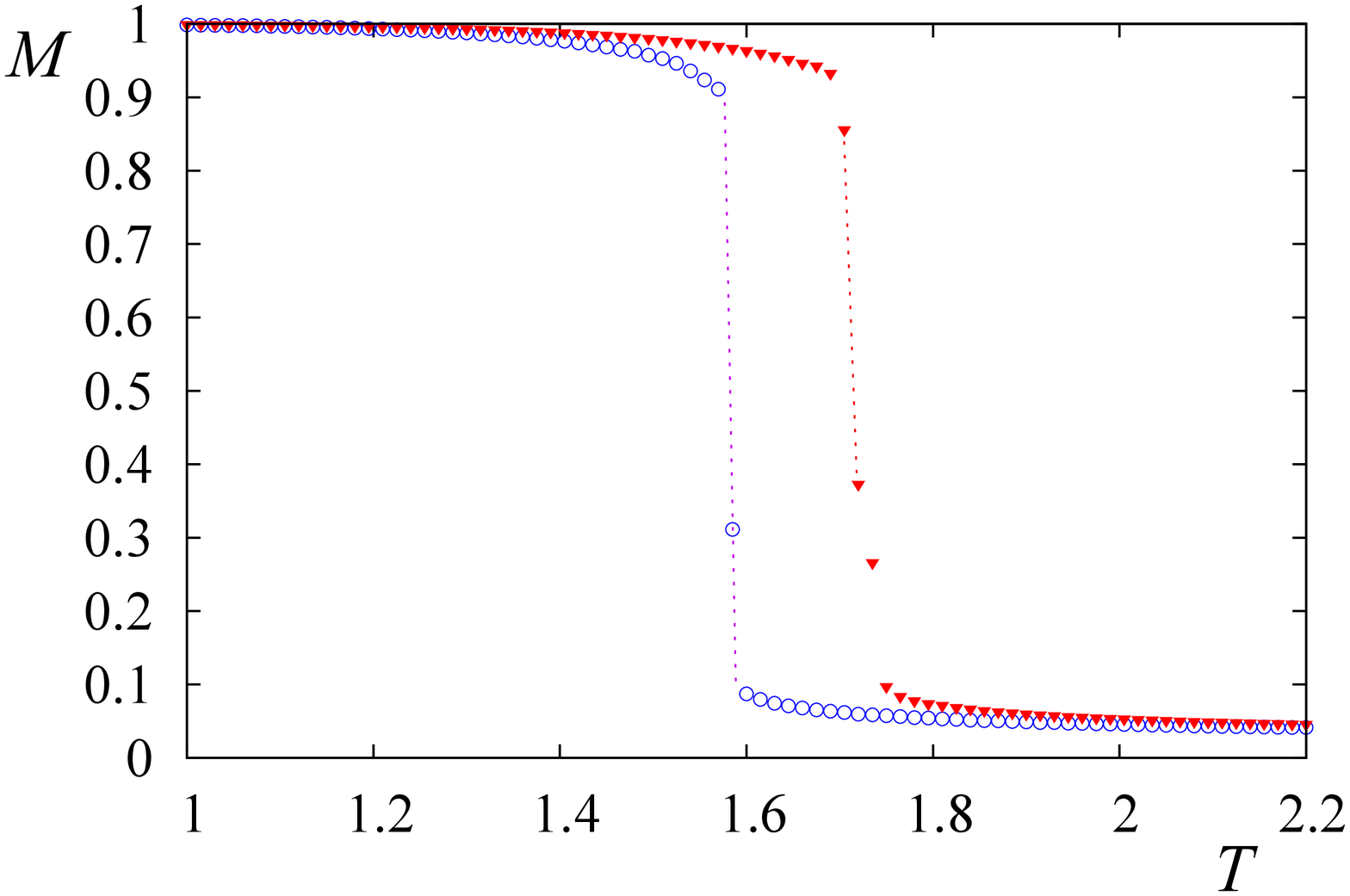}
\caption{Ising case  (color online): Energy $E$ and order parameter $M$ for  $\eta=J_1/J_2=0.85$ (blue void circles) and 1 (red triangles).   See text for comments.} \label{EMI}
\end{figure}

\begin{figure}
\centering
\includegraphics[width=50mm,angle=0]{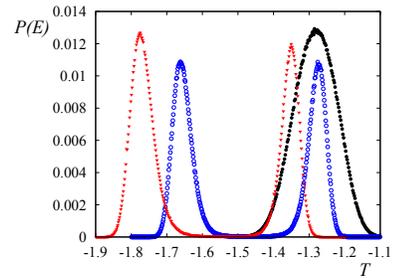}
\caption{Ising case (color online): Energy histogram $P(E)$ versus $E$ for $\eta=1$( red triangles), 0.85 (blue void circles) and 0.3 (black circles).   See text for comments.} \label{PI}
\end{figure}

In the following, the results  in different figures are shown with  $L=18$ and $L_z=8$ (16 atomic planes along $z$).
Since the GS changes at $\eta_c=0.5$, we show here examples on both sides of this value.
Figure \ref{EMI03} shows the energy per spin $E$, the specific heat per spin $C_V$, the order parameter $M$ and the susceptibility $\chi$, for $\eta=0.3$. The transition is of second order.   On the other side,
we show in Fig. \ref{EMI} the energy per spin and the order parameter versus $T$, for  $\eta=0.85$ and 1.   We find a strong first-order transition in both cases.  The discontinuity of $E$ and $M$ at the transition is very large.   We show in Fig. \ref{PI}
the energy histogram taken at the transition temperature for three values $\eta=0.3$ (black), 0.85 (blue) and 1 (red).   As seen, the first case is a gaussian distribution indicating a second-order transition, while the last two cases show a double-peak structure indicating a first-order transition.

\begin{figure}
\centering
\includegraphics[width=50mm,angle=0]{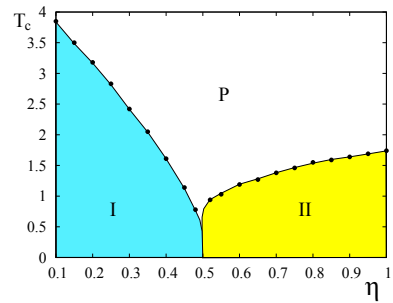}
\caption{Ising case: $T_C$ versus $\eta$. I, II and P denote the first, second and paramagnetic phases, respectively.   See text for comments.} \label{TcI}
\end{figure}

We have calculated the critical temperature $T_C$ as a function of $\eta$.  The phase diagram is shown in Fig. \ref{TcI} where I and II  indicate the ordering of the first, and second kinds, respectively. P indicates the paramagnetic phase.  Note that the transition line between I and P is a second-order line, while that between II and P is a first-order line.

\subsection{XY case}

\begin{figure}
\centering
\includegraphics[width=50mm,angle=0]{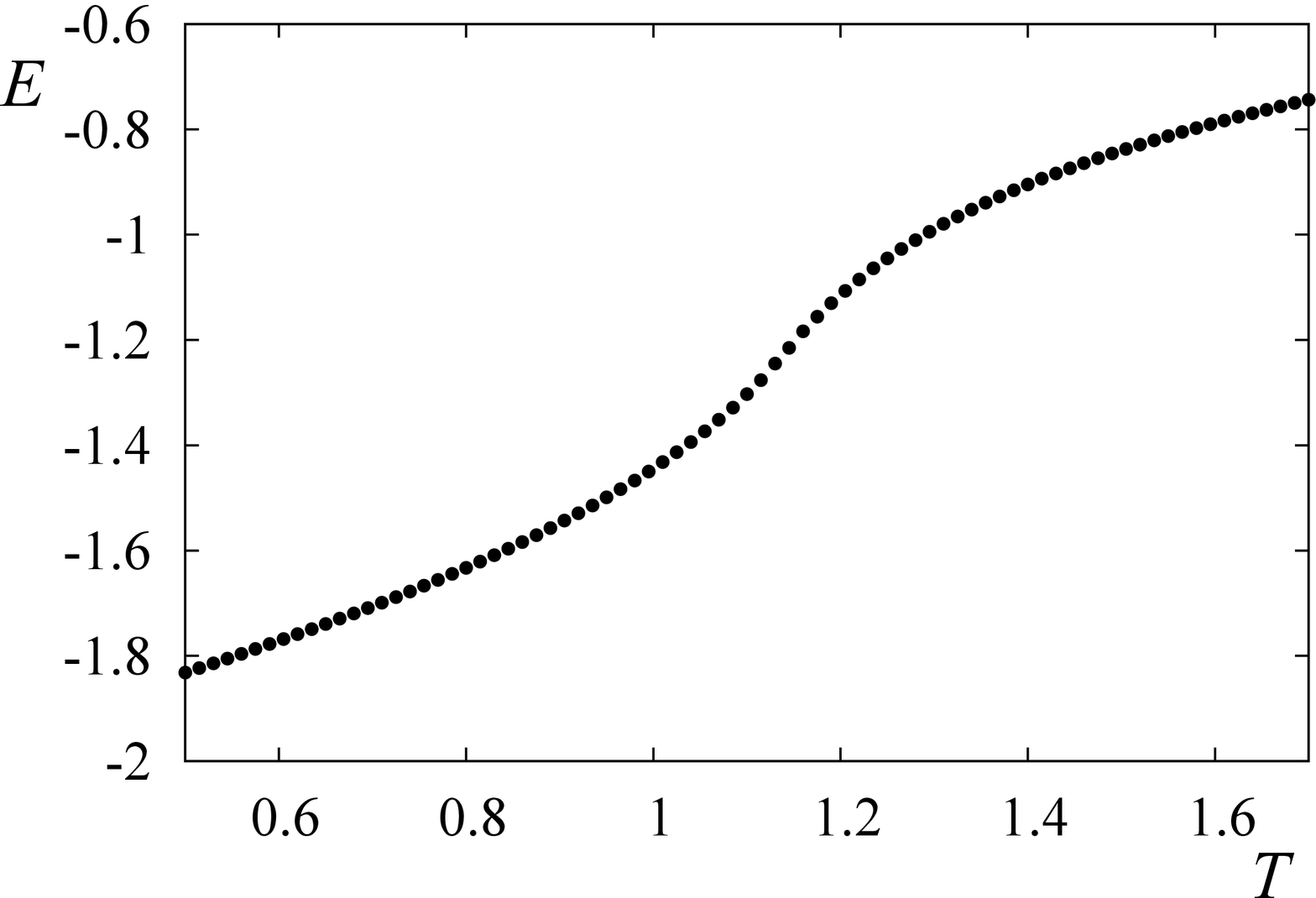}
\includegraphics[width=50mm,angle=0]{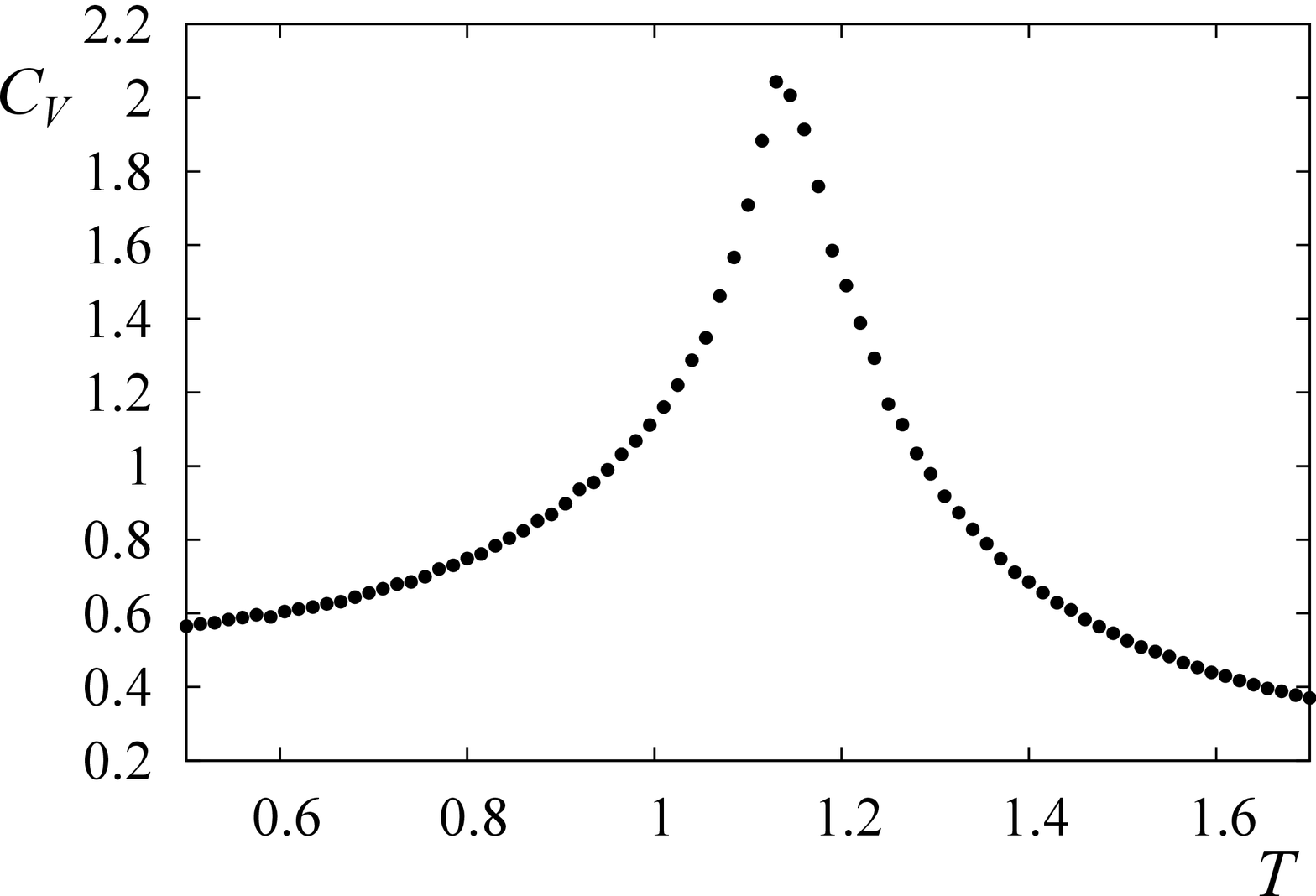}
\includegraphics[width=50mm,angle=0]{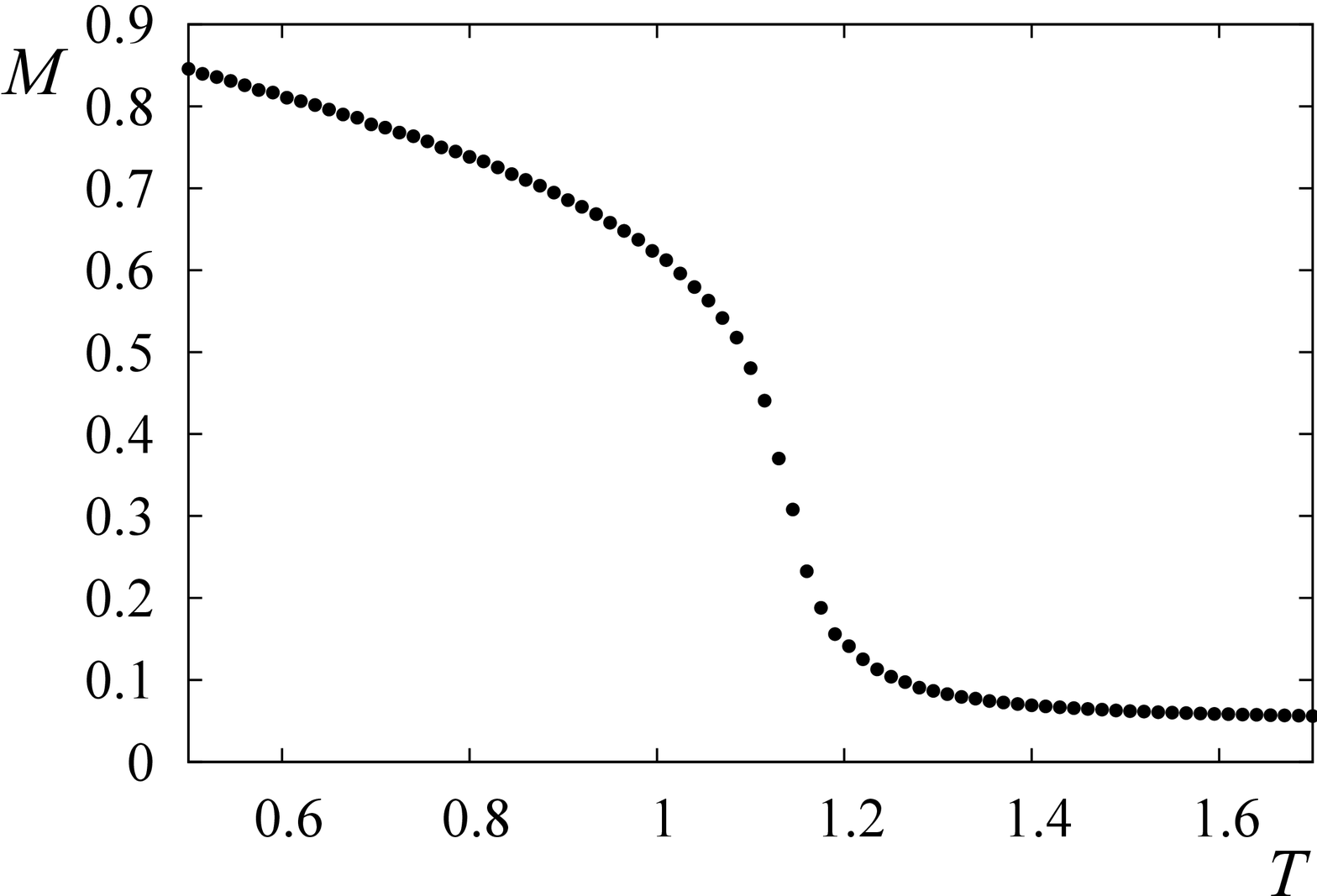}
\includegraphics[width=50mm,angle=0]{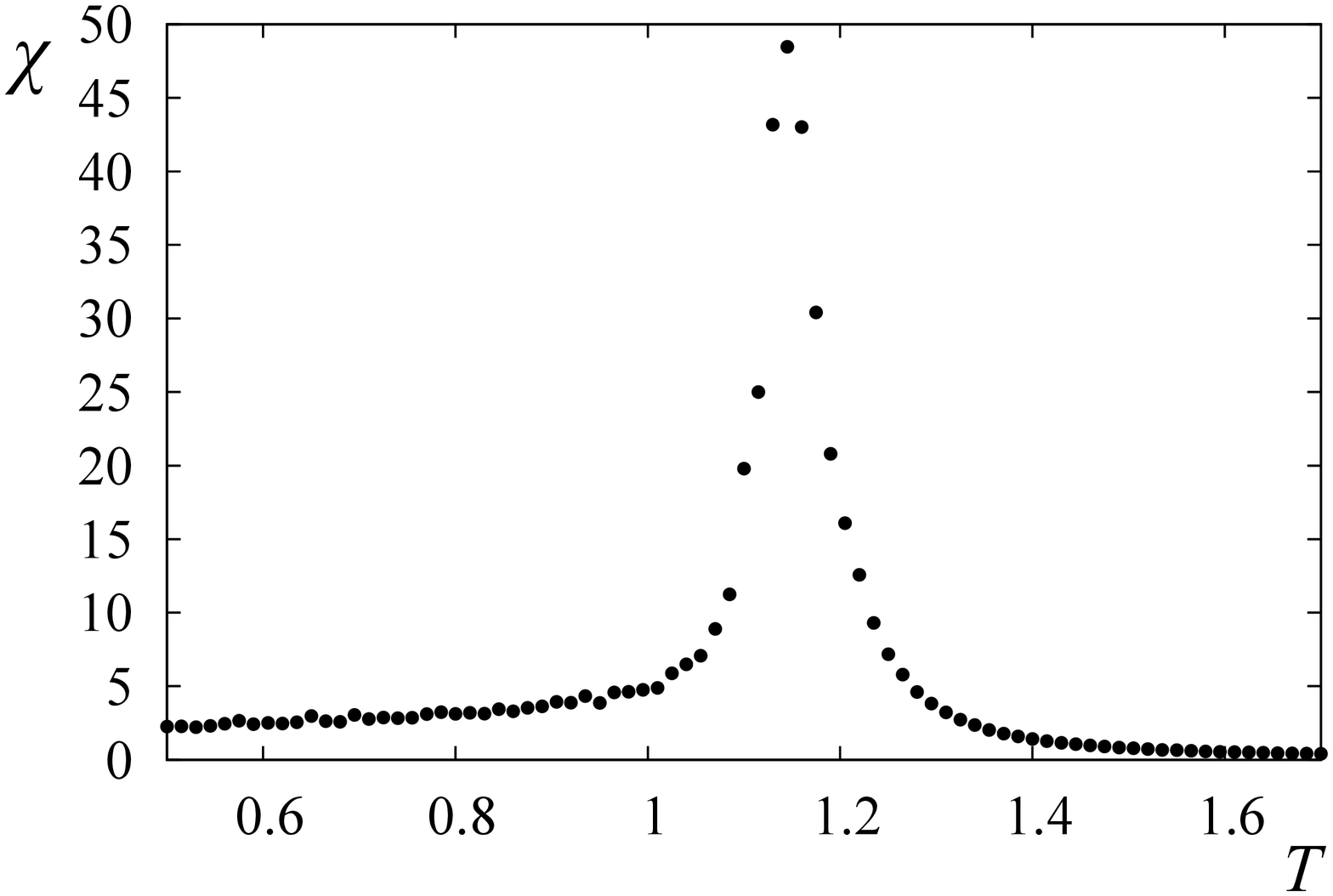}
\caption{XY case: Energy $E$, specific heat $C_V$, order parameter $M$ and susceptibility $\chi$  versus $T$  for $\eta=0.3$.   See text for comments.} \label{ECJ03XY}
\end{figure}

In the XY case, the change of the GS takes place at $\eta=1/3$.  Let us show in Fig. \ref{ECJ03XY} the result for $\eta=0.3$ where the GS is composed of ferromagnetic planes antiferromagnetically stacked in the $z$ direction.  The transition is of second order.

\begin{figure}
\centering
\includegraphics[width=50mm,angle=0]{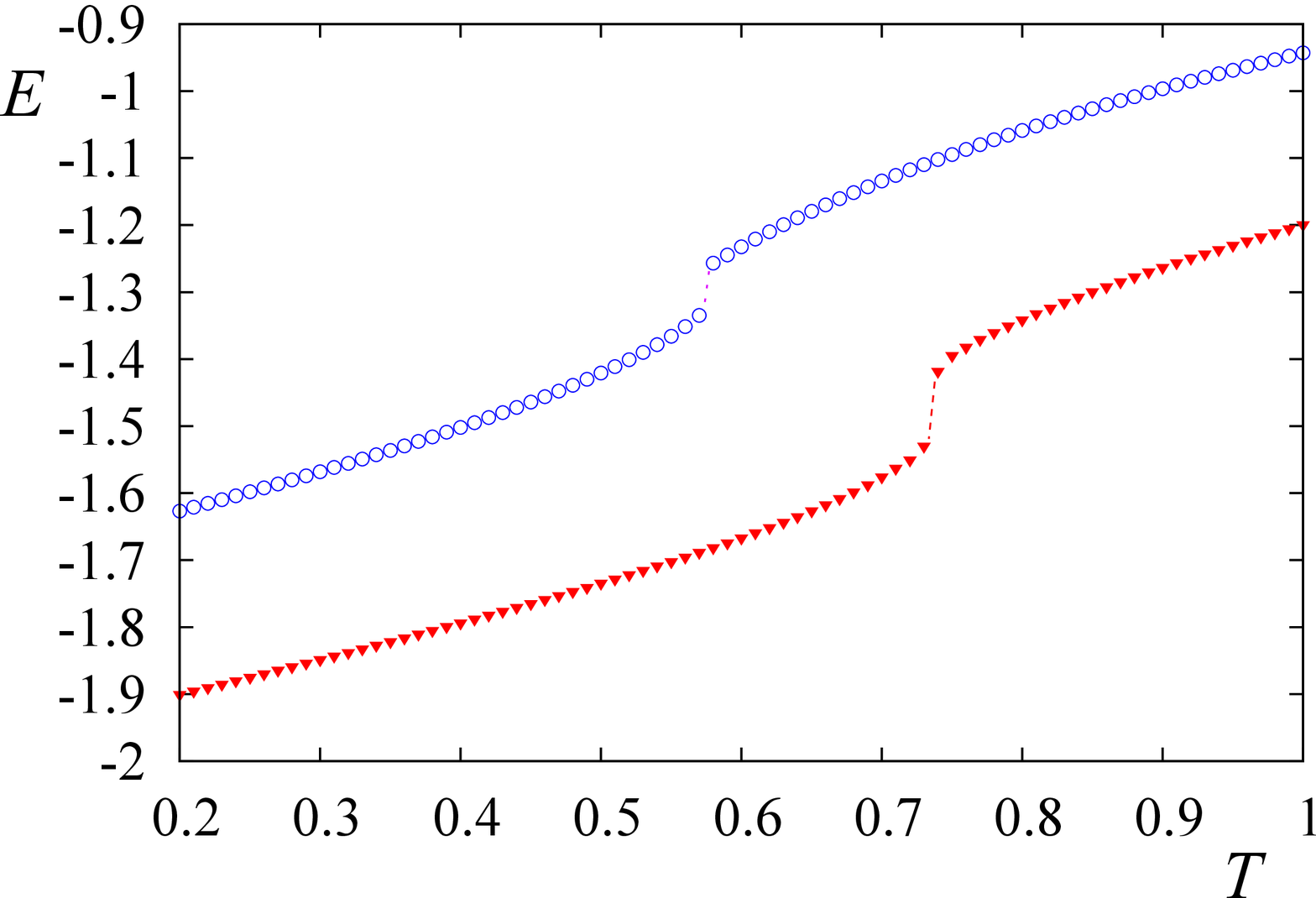}
\includegraphics[width=50mm,angle=0]{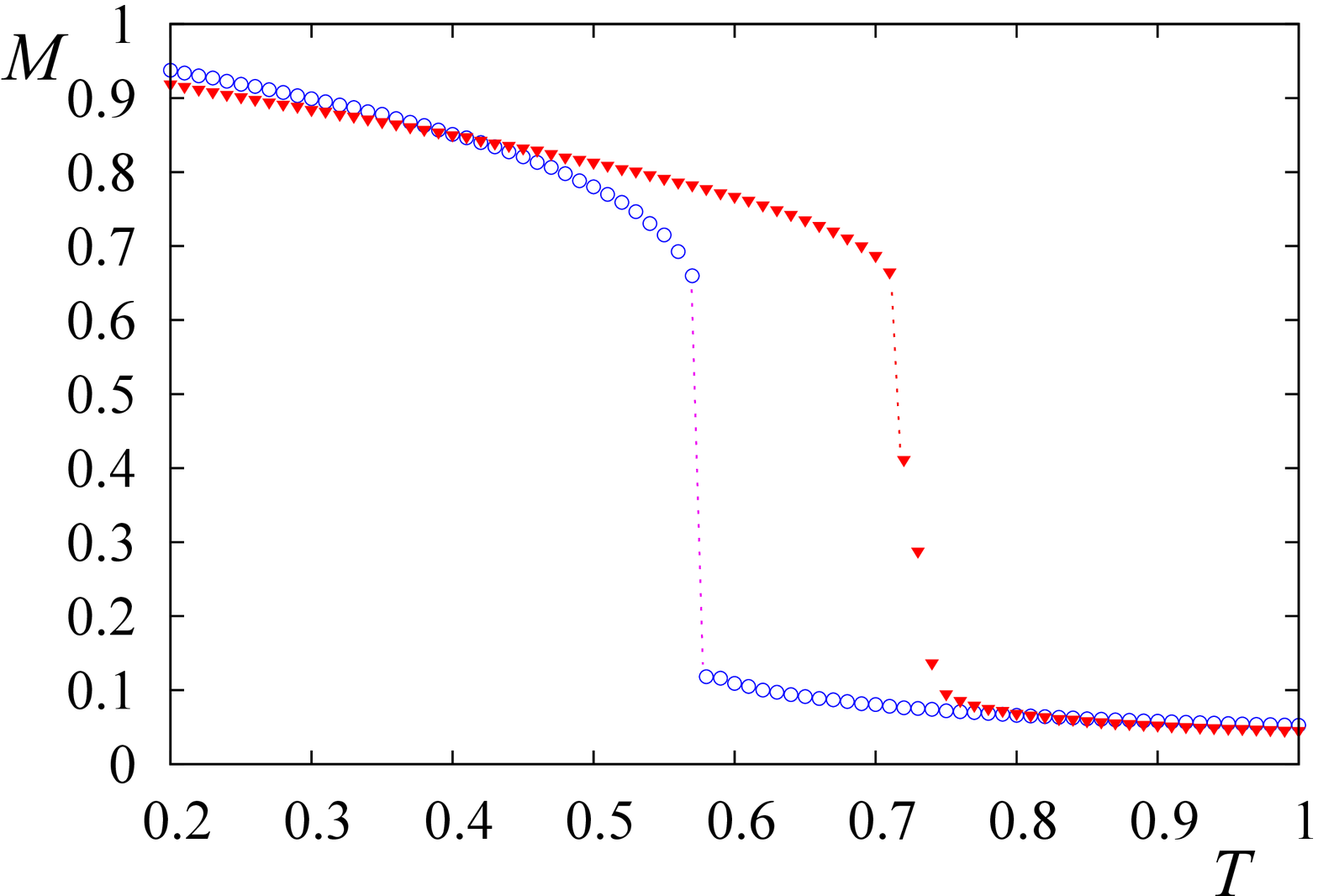}
\caption{XY case  (color online): $E$  and $M$ versus $T$ for $\eta=0.58$ (blue void circles) and 1 (red triangles).  See text for comments.} \label{EMJ058and1XY}
\end{figure}

\begin{figure}
\centering
\includegraphics[width=50mm,angle=0]{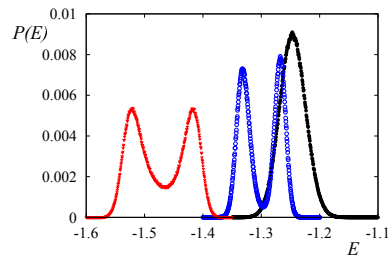}
\caption{XY case  (color online): Energy histogram $P$ versus $E$  for $\eta=0.3$ (black circles), 0.58 (blue void circles),  1 (red triangles) at the respective transition temperatures.  See text for comments.} \label{PXY}
\end{figure}

We show now the result for the non collinear GS region in Fig. \ref{EMJ058and1XY}. The energy and the order parameter show clearly a discontinuity at the transition for $\eta=0.58$ and 1.
Using the histogram method described above, we have calculated the histogram shown in Fig. \ref{PXY} for  $\eta=0.3$, 0.58 and 1.  For $\eta=0.3$ which is in the collinear region of the GS, the histogram is gaussian,  confirming the second-order transition observed in the data shown above.  For $\eta=0.58$ and 1 belonging to the non collinear region, the histogram shows a two-peak structure which confirms the  first-order character of the transitions in this region.

The two peaks are very well separated with the dip going down to zero, indicating  an energy discontinuity.  The distance between the two peaks is the latent heat $\Delta E$.

\begin{figure}
\centering
\includegraphics[width=50mm,angle=0]{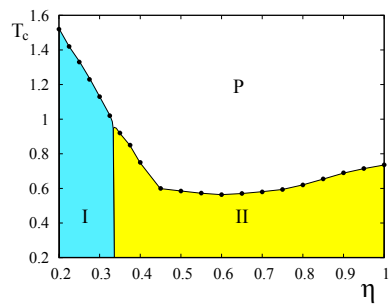}
\caption{XY case: $T_C$ versus $\eta$. I, II and P denote the collinear, non collinear and paramagnetic phases, respectively.  See text for comments.} \label{TcXY}
\end{figure}

We show in Fig. \ref{TcXY} the transition temperature versus $\eta$ where I and II indicate the collinear and non collinear phases, respectively. P denotes the paramagnetic state.   The line separating I and P is a second-order transition line, while that separating II and P is the first-order one.

To close this section, we emphasize that all 3D frustrated systems known so far undergo a first-order transition: let us mention the AF STL \cite{Ngo2008a,Ngo2008b}, the FCC antiferromagnets \cite{Diep-Kawamura}, the simple cubic fully frustrated lattices \cite{Nihat,Ngo2010,Ngo2011a,Ngo2011b},  helimagnets \cite{Diep1989}  and the HCP lattice studied here.

\section{Spin Transport in the Ising Case}

We have recently studied the spin transport in magnetically ordered materials by using an efficient MC simulation method \cite{Magnin,Magnin3}.  The tour-de-force of the method was realized when we  studied the semiconducting MnTe where the agreement with experiment is  excellent \cite{Magnin4}.  In ferromagnets, the spin resistivity shows a sharp peak at the magnetic phase transition.  The origin of this anomaly comes from the spin-spin correlation \cite{deGennes,Fisher,Kataoka}.  In antiferromagnets, such a peak is not sharp and often a broad maximum \cite{Haas}. We have also studied the spin resistivity in two frustrated cases: the FCC Ising case \cite{Magnin2} and the $J_1-J_2$ simple cubic lattice \cite{Hoang}.  The results show that the discontinuous transition in these systems gives rise to a discontinuity of the spin resistivity.

Using the method which has been described in detail elsewhere \cite{Magnin,Magnin2}, we calculate the current of itinerant spins moving  across the crystal under an electric field  $\vec \epsilon$ applied in the $x$ direction.  We suppose that each itinerant spin $\sigma_i$ interacts with the lattice spins around it within a sphere of radius $D_1$ at any point on its trajectory.  The interaction Hamiltonian reads

\begin{equation}
{\cal H}_l=- \sum_{j} I_{ij} \sigma_i\cdot S_j
\end{equation}
where the sum is performed over all lattice spins in the sphere around the itinerant spin $\sigma_i$ and $I_{ij}$ is the interaction which depends on the distance between $\sigma_i$ and $S_j$.  We suppose also the following interaction between an itinerant spin with its neighboring itinerant spins within a sphere of radius $D_2$

\begin{equation}
{\cal H}_i=- \sum_{j} K_{ij} \sigma_i\cdot \sigma_j
\end{equation}
For simplicity, we suppose
\begin{eqnarray}
I_{ij}&=&I_0 \exp(-Br_{ij}) \\
K_{ij}&=&K_0 \exp(-Cr_{ij}) \label{K}
\end{eqnarray}
where $I_0$, $B$, $K_0$ and $C$ are constants chosen in such a way that the energy of an itinerant spin
is smaller than that of a lattice spin. This choice is made to avoid the influence
of itinerant spins on the ordering of the lattice spins.  Note that in the almost-free
electron model, $I_0\simeq K_0\simeq 0$, and in semiconductors they are larger but still weak with
respect to $J_1$ and $J_2$.  A discussion on the choice of different parameters has been given for example
in Ref.\cite{Magnin}.  We suppose here  a concentration of one
itinerant per two lattice cells.   At such a low concentration, the averaged distance between itinerant spins is much larger than the cutoff distance $D_2$.  In spite of this, due to the attractive nature of their interaction, Eq. (\ref{K}), we have to use a chemical potential term to insure that itinerant spins do not form  clusters. This potential is taken of the form $D[n(\vec r)-n_0]$ where $D$ is a positive constant, $n(\vec r)$ the concentration of itinerant spins in the sphere of  cutoff radius $D_2$ centered at the position $\vec r$ of the itinerant spin under consideration, and $n_0$ the averaged concentration.  We perform the simulation by taking into account the relaxation time of the lattice spins \cite{Magnin3,Magnin4,Hohenberg} which is given by

\begin{equation}\label{tau}
\tau_L=\frac{A}{|1-T/T_C|^{z\nu}}
\end{equation}
where $A$ is a constant, $\nu$  the correlation critical exponent, and $z$ the dynamic exponent.  We see that as $T$ tends to $T_C$, $\tau_L$ diverges.  This phenomenon is known as the critical slowing-down.  For the Ising spin model, $\nu=0.638$ (3D Ising universality) and $z=2.02$ \cite{Prudnikov}.  We have previously shown  that $\tau_L$ strongly  affects the shape of  $\rho$ at $T_C$ \cite{Magnin3}.  By choosing $A=1$, we fix $\tau_L=1$ at $T=2T_C$ deep inside the paramagnetic phase far above $T_C$. This value is what we expect for thermal fluctuations in the disordered phase.

\begin{figure}
\centering
\includegraphics[width=60mm,angle=0]{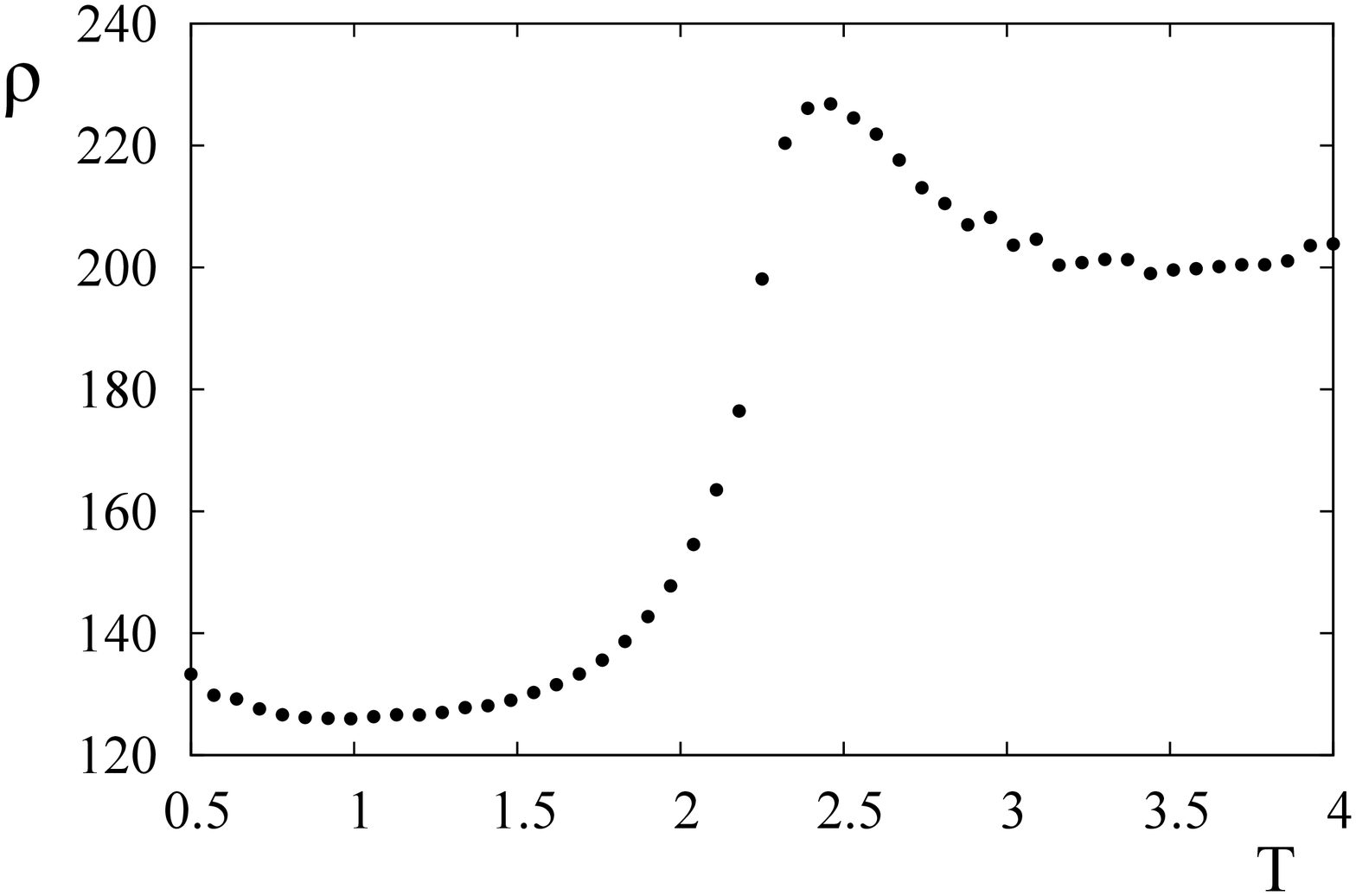}
\includegraphics[width=60mm,angle=0]{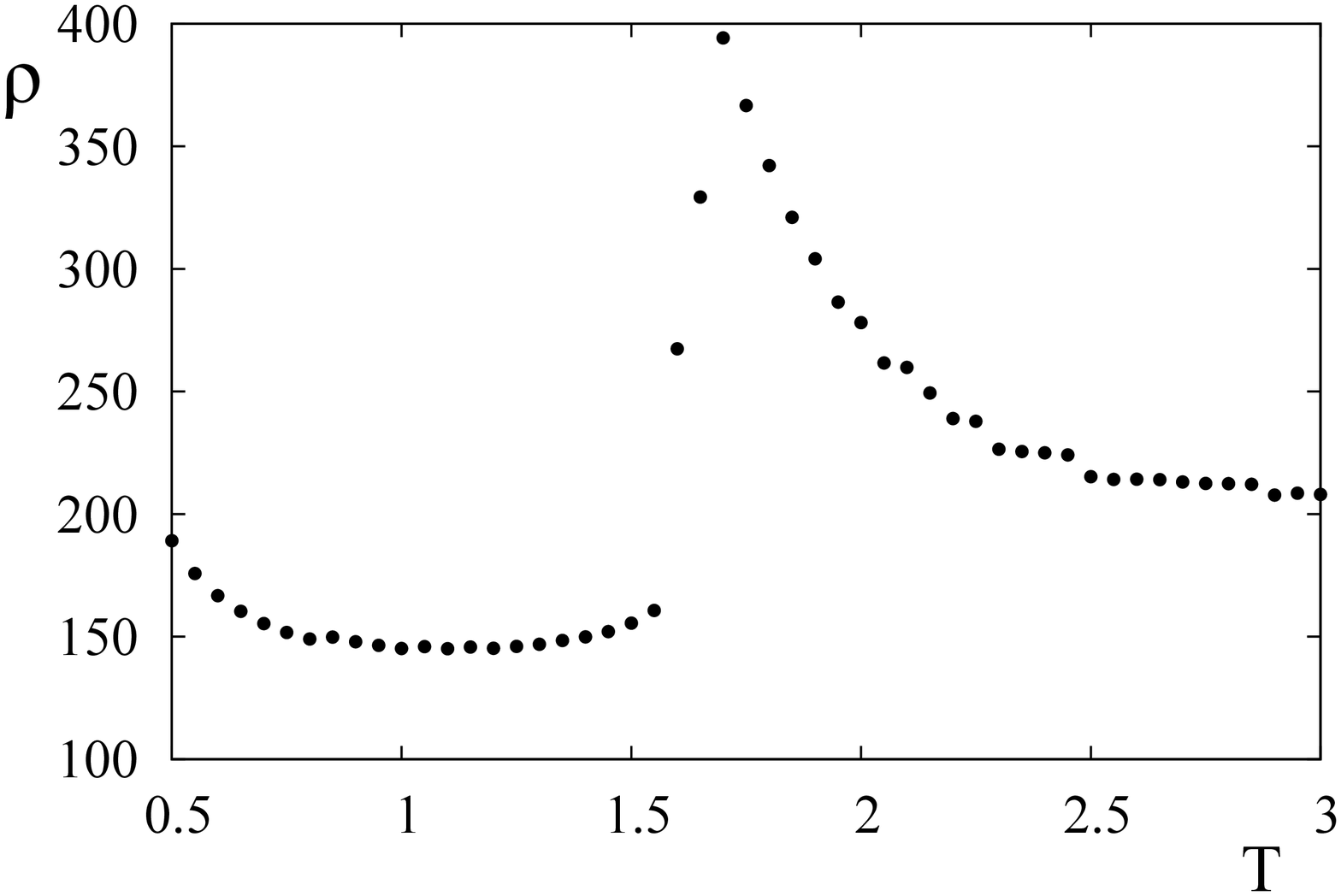}
\caption{Ising case: Spin resistivity $\rho$ versus $T$ for $\eta=0.3$ (upper) and 1 (lower).   $L=18$, $L_z=8$ (16 planes in the $z$ direction), $D_1=D_2=1$, $\epsilon=1$, $I_0=2$, $K_0=0.5$, $B=C=1$, $A=1$, $D=0.5$.  All distances are in unit of the NN distance, energy constants are in unit of $|J_2|=1$.  See text for comments.} \label{R1}
\end{figure}

The spin resistivity $\rho$ versus $T$ for the two typical cases $\eta=0.3$ and 1
is shown in Fig. \ref{R1} where distances $D_1$ and $D_2$ are in unit of the NN distance, energy constants $I_0$, $K_0$ and $D$ are in unit of $|J_2|=1$.  We observe here that in the second-order region $\rho$  has
a  rounded maximum and  in the first-order region it undergoes an almost discontinuous jump at the transition.

We have varied the radius $D_1$ to see its effect on the resistivity $\rho$ at the transition.  We observe the same effect which has been seen  in some other   antiferromagnets \cite{Magnin2,Hoang}: at a given $T$, $\rho$
oscillates slightly with distance. We have found that this oscillation comes from the  oscillatory behavior of the difference between the numbers of up and down spins in the sphere as $D_1$ varies.


 \section{Conclusion}\label{conclu}

 We have studied in this paper some properties of the HCP antiferromagnet with Ising and XY spin models. The in-plane $J_1$ and inter-plane $J_2$ interactions are supposed to be different. As a result, the GS spin configuration depends on the ratio $\eta=J_1/J_2$. We show that there exists a critical value $\eta_c$ where the GS changes. For the Ising case, we find $\eta_c=0.5$ below (above) which the spins in the  $xy$ planes are ferromagnetic (antiferromagnetic). For the XY case,  the GS is collinear below $\eta_c=1/3$, and is  non collinear above that value.   The nature of the transition changes from a second order below $\eta_c$ to a first order above $\eta_c$ for both Ising and XY cases.    We have also studied the spin resistivity in the Ising case. We found that the shape of $\rho$ depends on the nature of the transition: in the second-order region, a rounded maximum is observed at the transition while in the first-order region, $\rho$ undergoes a discontinuity at the transition as we have observed in other frustrated cases. These findings may help to understand the transition nature and the spin transport in different compounds with HCP structure.

{}

\end{document}